\documentclass[aps,pr,twocolumn,showpacs,superscriptaddress]{revtex4}
\usepackage{epstopdf}
\usepackage[colorlinks]{hyperref}
\usepackage{amssymb}
\usepackage{color}
\usepackage{graphicx,amsmath}
\hypersetup{colorlinks,citecolor=red,linkcolor=blue,urlcolor=blue}

\begin{document}

\title{Fermion condensation, $T$-linear resistivity and Planckian limit}
\author{V. R. Shaginyan}
\email{vrshag@thd.pnpi.spb.ru}\affiliation{Petersburg Nuclear
Physics Institute of NRC "Kurchatov Institute", Gatchina,
188300, Russia}\affiliation{Clark Atlanta University, Atlanta,
GA 30314, USA} \author{M. Ya. Amusia}\affiliation{Racah
Institute of Physics, Hebrew University, Jerusalem 91904,
Israel}\affiliation{Ioffe Physical Technical Institute, RAS, St.
Petersburg 194021, Russia} \author{A. Z.
Msezane}\affiliation{Clark Atlanta University, Atlanta, GA
30314, USA}
\author{V. A. Stephanovich}\affiliation{Institute of Physics, Opole
University, Oleska 48, 45-052, Opole, Poland}\author{G. S.
Japaridze}\affiliation{Clark Atlanta University, Atlanta, GA
30314, USA}\author{S. A. Artamonov} \affiliation{Petersburg
Nuclear Physics Institute of NRC "Kurchatov Institute",
Gatchina, 188300, Russia}

\begin{abstract}
We explain recent challenging experimental observations of
universal scattering rate related to the linear-temperature
resistivity exhibited by a large corps of both strongly
correlated Fermi systems and conventional metals. We show that
the observed scattering rate in strongly correlated Fermi
systems like heavy fermion metals and high-$T_c$ superconductors
stems from phonon contribution that induce the linear
temperature dependence of a resistivity. The above phonons are
formed by the presence of flat band, resulting from the
topological fermion condensation quantum phase transition
(FCQPT). We emphasize that so - called Planckian limit, widely
used to explain the above universal scattering rate, may occur
accidentally as in conventional metals its experimental
manifestations (e.g. scattering rate at room and higher
temperatures) are indistinguishable from those generated by the
well-know phonons being the classic lattice excitations. Our
results are in good agreement with experimental data and show
convincingly that the topological FCQPT can be viewed as the
universal agent explaining the very unusual physics of strongly
correlated Fermi systems.
\end{abstract}
\maketitle

Exotic experimentally observable properties of different classes
of strongly correlated Fermi systems are still remain largely
unexplained due to the lack of universal underlying physical
mechanism. It is customary to attribute these properties to
so-called non-Fermi-liquid (NFL) behavior. Latter behavior is
widely observed in heavy-fermion (HF) metals, graphene, and
high-$T_c$ superconductors (HTSC). Experimental data collected
on many of these systems show that at $T=0$ a portion of their
excitation spectrum becomes dispersionless, giving rise to
so-called flat bands, see e. g. \cite{ks,vol,graph,noz}. The
presence of flat band indicates that the system is close to a
special quantum critical point (QCP), in which the topological
class of Fermi surface alters. This QCP is coined as topological
fermion-condensation quantum phase transition (FCQPT)
\cite{ks,vol,graph}, leading to flat bands ($\varepsilon ({\bf
k})=\mu$, where  $\varepsilon ({\bf k})$ is quasiparticle energy
and $\mu$ is a chemical potential) formation. Latter phenomenon
is called fermion condensation (FC) and had been predicted long
ago \cite{ks,vol}. The flat bands are formed by interaction
(while a {{geometric}} frustration can help the process), and
have a special $T-$ dependence: At rising temperatures, {{a
nonzero slope appears in the $\varepsilon ({\bf k})$ dispersion
law}}, while the quasiparticle width $\gamma\propto T$
\cite{khod95,phys_rep}. This observation is in accordance with
experimental data, see e. g. \cite{meln17,liu18,shash19}.
Moreover, the FC theory allows one {{to describe adequately
(both qualitatively and quantitatively) the above}} NFL behavior
of strongly correlated Fermi systems
\cite{ks,vol,prr1,phys_rep,book,vol2018,vol2019}. {{Here we
analyze HF metals and high-$T_c$ superconductors}} exhibiting
$\rho(T)\propto T$ at $T\to0$ and, therefore, located near their
topological FCQPT \cite{book,quasi}. {{Note that FC theory
permits also to consider the systems located relatively far from
their FCQPTs, see e. g. \cite{phys_rep,book}.

Recent challenging experimental findings of linear temperature
dependence of the resistivity $\rho(T)\propto T$ collected on
HTSC, graphene, HF and conventional metals, have revealed that
the scattering rate 1$/\tau$ of charge carriers reaches the
so-called universal Planckian limit $1/(T\tau)=k_B/\hbar$ ($k_B$
and $\hbar=h/2\pi$ are the Boltzmann and Plank constants,
respectively) \cite{bruin,legr,cao,nakaj}. Note that above
Planckian limit, used to explain the universal scattering rate
in the so called Planckian metals
\cite{bruin,legr,cao,nakaj,sach19}, can occur accidentally since
its experimental manifestations in other (than Planckian metals)
metals may be equally well explained by more conventional
physical mechanisms like those related to phonon contribution
\cite{quasi,vol2019}. For instance, the conventional metals
exhibit the universal linear scattering rate at room and higher
temperatures, generated by the well-known phonons being the
classic lattice excitations  \cite{bruin}.}}

In the present Letter, within the framework of the FC theory, we
show that the quasi-classical physics is still applicable to
describe the universal scattering rate 1$/\tau$ experimentally
observed in strongly correlated metals at their quantum critical
region. This is because flat bands, responsible for quantum
criticality, generate transverse zero-sound mode, reminiscent of
the phonon mode in solids, with the Debye temperature $T_D$
\cite{quasi,DP,vol2019}. At $T\geq T_D$ the mechanism of the
$T$-linear dependence of the resistivity is the same both in
conventional metals and strongly correlated ones, and is
represented by electron-phonon scattering. Therefore, it is
electron-phonon scattering at $T\geq T_D$ that leads to the near
material-independence of the lifetime $\tau$ that is expressed
as $1/(\tau T)\sim k_B/\hbar$. As a result, we describe and
explain recent exciting experimental observations of universal
scattering rate related to linear-temperature resistivity of a
large number of both strongly correlated Fermi systems and
conventional metals \cite{bruin,legr,cao,nakaj}. Thus, the
observed scattering rate is explained by the emergence of flat
bands formed by the topological FQCPT, rather than by the so
called Planckian limit at which the assumed Planckian scattering
rate takes place. {{The reason is that Planckian limit occurs in
the conventional metals at relatively high temperatures $T>T_D$,
while it is normally takes place at low temperatures $T\to0$,
but it does not \cite{bruin}.}}

\begin{figure} [! ht]
\begin{center}
\includegraphics [width=0.47\textwidth]{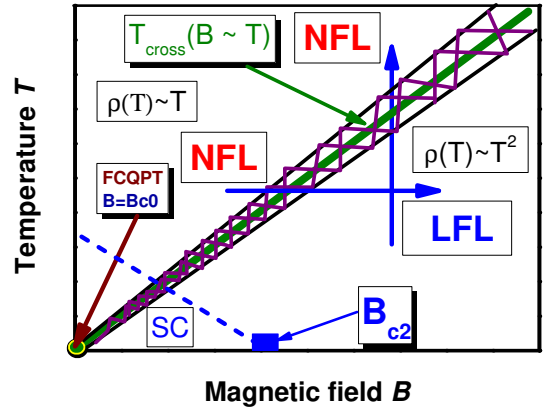}
\end{center}
\caption{(color online). Schematic $T-B$ phase diagram of a
strongly correlated Fermi system. The vertical and horizontal
arrows crossing the transition region marked by the thick lines
depict the LFL-NFL and NFL-LFL transitions at fixed $B$ and $T$,
respectively.  {{At $B<B_{c2}$ (dashed line beginning at solid
rectangle) the system is in its possible SC state, with $B_{c2}$
being a critical magnetic field. The hatched area with the solid
curve $T_{\rm cross}(B\sim T)$ represents the crossover
separating the NFL and LFL domains. A part of the crossover is
hidden inside possible SC state. The boxes $\rho(T)\propto T$
and $\rho(T)\propto T^2$ demonstrate the NFL and LFL behavior of
resistivity, respectively.}} }\label{Fig2PD}
\end{figure}

We begin with considering the schematic $T-B$ phase diagram of
strongly correlated Fermi system \cite{epl14} depicted in Fig.
\ref{Fig2PD}. The magnetic field $B$ plays a role of the control
parameter, driving the system towards its QCP represented by
FCQPT. The FCQPT occurs at $B=B_{c0}$, yielding new strongly
degenerate state at $B=B_{c0}$. To lift this degeneracy, the
system forms either superconducting (SC), magnetically ordered
(ferromagnetic (FM), antiferromagnetic (AFM) etc.), or nematic
states \cite{phys_rep}. In the case of $\rm CeCoIn_5$, this
state is located at $B_{c0}\simeq 4.9$ T and covered with
superconducting "dome" with the critical field $B=B_{c2}\simeq
5$ T \cite{pag1}. In case of $\rm Sr_3Ru_2O_7$ $B_{c0}\simeq
7.9$ T, while the SC state is absent \cite{bruin}. It is seen
from Fig. \ref{Fig2PD}, that at fixed temperature the increase
of $B$ drives the system along the horizontal arrow from NFL
state to LFL one. On the contrary, at fixed magnetic field and
rising temperatures the system transits along the vertical arrow
from LFL state to NFL one. The region shown by the arrow and
labeled by $T_{\text{cross}}(B\sim T)$ signifies a transition
regime between the LFL part with almost constant effective mass
and NFL one at which $\rho(T)\propto T$.

\begin{figure} [! ht]
\begin{center}
\includegraphics [width=0.47\textwidth] {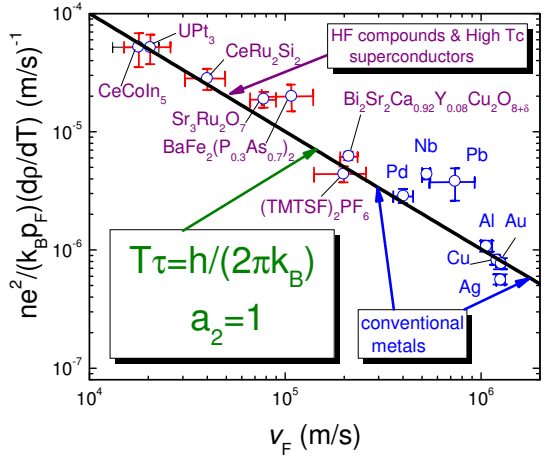}
\end {center}
\caption{(color online). Scattering rates per kelvin of
{{different strongly correlated metals like HF metals, HTSC,
organic and conventional metals}} \cite{bruin}. All these metals
exhibit $\rho(T)\propto T$ and demonstrate two orders of
magnitude variations in their Fermi velocities $v_F$. The
parameter $a_2\simeq 1$ gives the best fit {{shown by the solid
line, and corresponds to the scattering rate $\tau T=h/(2\pi
k_B)$, with $h=2\pi \hbar$, see Eqs. \eqref{vf} and
\eqref{planc}.}} The region occupied by the conventional metals
is {{displayed by the two arrows. The arrow shows the region of
strongly correlated metals, including organic ones.}} Note, that
at low temperatures $T\ll T_D$, the scattering rate per kelvin
of a conventional metal is orders of magnitude lower, and does
not correspond to the Planckian limit.} \label{Sc1}
\end{figure}

{{At low temperatures, the observed  resistivity in HTSC and HF
metals located near their QCPs, obeys linear law (so-called
linear $T$-resistivity)}}
\begin{equation}
\rho(T)=\rho_0+AT.\label{res}
\end{equation}
{{This law demonstrates their quantum criticality and new state
of matter \cite{quasi,jltp:2017}.}} Here $\rho_0$ is the
residual resistivity and $A$ is a $T$-independent coefficient.
Explanations based on quantum criticality for the $T$-linear
resistivity have been given in the literature, see e. g.
\cite{lee-h,varma,varma1,phill,phill1,DP,quasi,khod2012} and
{{references therein. On the other hand, at room temperature
the}} $T$-linear resistivity is exhibited by conventional metals
such as $\rm Al$, $\rm Ag$ or $\rm Cu$. In case of a simple
metal with a single Fermi surface pocket the resistivity reads
$e^2n\rho=p_F/(\tau v_F)$, \cite{trio} where $e$ is the
electronic charge, $\tau$ is the lifetime, $n$ is the carrier
concentration, and $p_F$ and $v_F$ {{are the Fermi momentum and
velocity respectively. Representing}} the lifetime $\tau$ (or
inverse scattering rate) of quasiparticles in the form
\cite{tomph,khod2012}
\begin{equation}\label{LT}
\frac{\hbar}{\tau}\simeq a_1+\frac{k_BT}{a_2},
\end{equation}
we obtain \cite{quasi}
\begin{equation}\label{vf}
a_2\frac{e^2n\hbar}{p_Fk_B}\frac{\partial\rho}{\partial
T}=\frac{1}{v_F},
\end{equation} where $a_1$ and $a_2$ are $T$-independent
parameters. A challenging point for a theory is that
experimental facts corroborate Eq. \eqref{vf} in case of both
strongly correlated metals (HF metals and HTSC) and ordinary
ones, provided that these demonstrate the linear $T$-dependence
of their resistivity \cite{bruin}, see Fig. \ref{Sc1}.

Moreover, {{the analysis of literature data for the various
compounds}} with the linear dependence of $\rho(T)$ shows: The
coefficient $a_2$ is always close to unity, $0.7\leq a_2\leq
2.7$, notwithstanding huge distinction in the absolute value of
$\rho$, $T$ and Fermi velocities $v_F$, varying by two orders of
magnitude \cite{bruin}. As a result, it follows from Eq.
\eqref{LT} that the $T$-linear scattering rate is of universal
form, $1/(\tau T)\sim k_B/\hbar$, regardless of different
systems displaying the $T$-linear dependence with parameter
entering Eq. \eqref{vf}, $a_2\simeq 1$, \cite{bruin,quasi,book}.
Indeed, this dependence is demonstrated by ordinary metals at
temperatures higher than the Debye {{one}}, $T\geq T_D$, with an
electron - phonon mechanism and by strongly correlated metals
which are assumed to be fundamentally different from the
ordinary ones, in which the linear dependence at their quantum
criticality and temperatures of a few Kelvin is assumed to come
from excitations of electronic origin rather than from phonons
\cite{bruin}. We note that in some of the cuprates the
scattering rate has a momentum and doping dependence omitted in
Eq. \eqref{vf} \cite{peets,french,alld}. As it is seen from Fig.
\ref{Sc1}, this scaling relation spans two orders of magnitude
in $v_F$, attesting to the robustness of the observed empirical
law \cite{bruin}. This behavior is explained within the
framework of the FC theory, since {{both for conventional metals
and strongly correlated ones the scattering rate is defined}} by
phonons \cite{quasi}. In case of conventional metals at $T>T_D$
it is well known that phonons make the main contribution to the
linear dependence of the resistivity, see e. g. \cite{trio}. On
the other hand, it has been shown that the quasi-classical
physics describes the $T$-linear dependence of the resistivity
of strongly correlated metals at $T>T_D$, since flat bands,
forming the quantum criticality, generate transverse zero-sound
mode with the Debye temperature $T_D$ located within the quantum
criticality area \cite{DP,khod2012,quasi}. Therefore, the
$T$-linear dependence is formed by electron-phonon scattering in
both ordinary metals and strongly correlated ones. As a result,
it is electron-phonon scattering that leads to the near
material-independence of the lifetime $\tau$ that is expressed
as
\begin{equation}\label{planc}
 \tau T\sim \frac{\hbar}{k_B}.
\end{equation}

Now we turn to the twisted bilayer graphene exhibiting the
universal scattering rate \cite{legr} and having flat band
\cite{graph}. Recent calculations have shown that under the
application of pressure the graphene produces increasing
correlated behavior, identified by the presence of flat bands at
twist angles that increase with growing pressure
\cite{carr2018}. Such a behavior signals that it is the
correlations that induce the flat bands, and is in accordance
with the scenario of FC \cite{ks,phys_rep}. We can qualitatively
figure out this observation: The twisted bilayer graphene can be
represented by a quasicrystal, that is long-range ordered and
yet non-periodic. In that case both the flat band and the
corresponding NFL behavior emerge \cite{shaginyan2013,book}. In
fact, without interlayer coupling, a monolayer twisted graphene
is not a quasicrystal, for it remains a periodic system, while
it is the coupling that makes the quasicrystal, and the
application of pressure strengthens the coupling. To support the
scenario, we predict that under the application of magnetic
field the observed $T-$linear dependence of the resistivity is
changed to the $T^2$ one, for the system transits from the NFL
state to LFL one as it is seen from phase diagram \ref{Fig2PD}
and the universality of the scattering rate is violated. In the
same way, the asymmetrical tunneling conductivity vanishes in
the LFL state, as it was predicted within the framework of the
FC theory \cite{shag2005,phys_rep,shag2018,ph_scr}. As a result,
we can safely surmise that it is the transverse sound mode that
forms the universal scattering rate rather than the Planckian
limit does.

\begin{figure}[!ht]
\begin{center}
\includegraphics[width=0.47\textwidth]{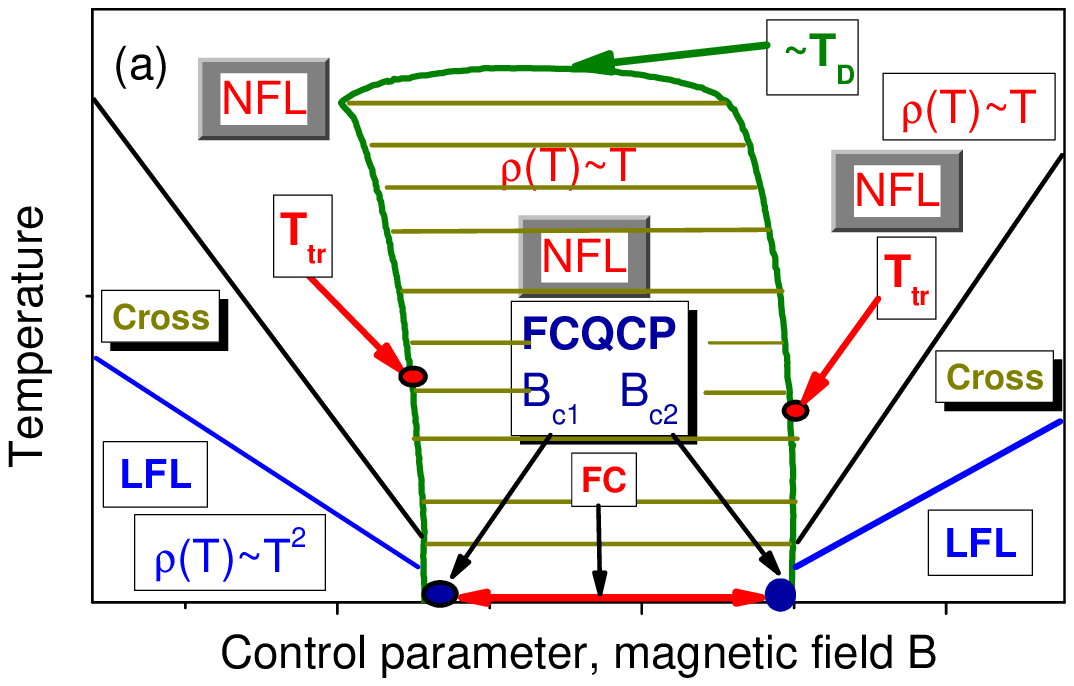}
\includegraphics[width=0.47\textwidth]{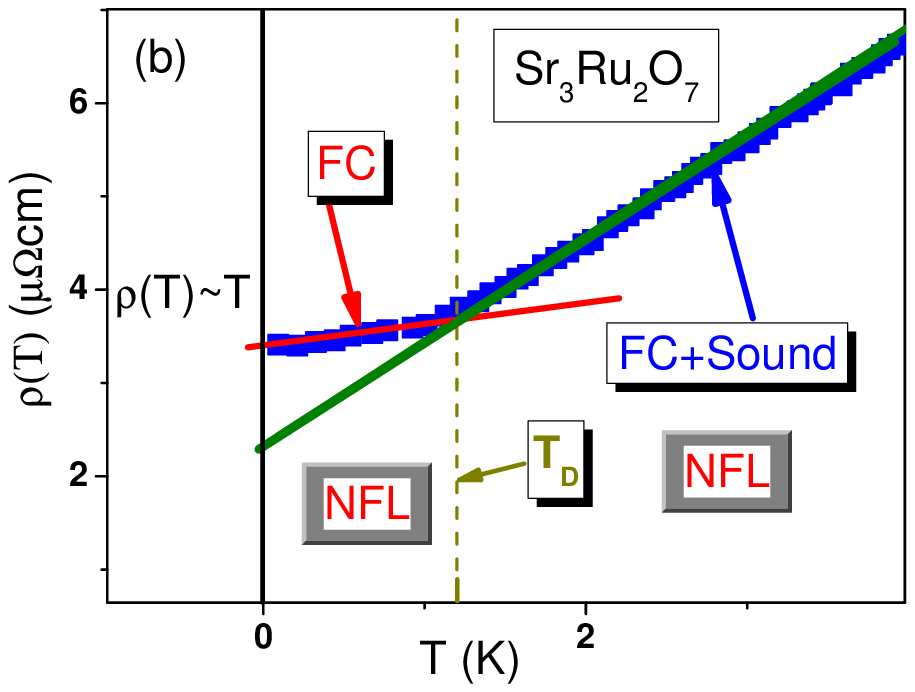}
\end{center}
\caption{(color online). Flat band formed by FC in the HF metal.
Panel (a). Schematic phase diagram of the metal $\rm
Sr_3Ru_2O_7$ based on experimental observations (see
\cite{bruin,pnas,sc9,physc} and references therein). The
topological FCQPTs located at magnetic fields $B$ within the
critical magnetic fields $B_{c1}\leq B\leq B_{c2}$ are indicated
by the arrows. The ordered nematic phase
\cite{bruin,pnas,sc9,physc} bounded by the thick curve and
demarcated by the horizontal lines emerges to remove the entropy
excess $S_0$ that emerges at $T<T_D$. The nematic phase sets in
at $T\simeq T_D$ Two arrows label the tricritical points
$T^1_{\rm tr}$ and $T^2_{\rm tr}$ at which the lines of
second-order phase transitions change to the first order. The
NFL behavior with the resistivity $\rho(T)\propto T$ induced by
both FC and transverse sound is labeled by NFL. The LFL behavior
with $\rho(T)\propto T^2$ is marked by LFL and shown in Fig.
\ref{Fig2} by the label, while the crossover regions are shown
by "Cross". Panel (b). {{The linear temperature dependence of
the resistivity, $\rho(T)\propto T$ at the NFL region.}} $T_D$
is the Debye temperature, marking the transition from
quasiclassic region to the nematic state, shown in panel (a) by
horizontal lines. The data are extracted from Ref.
\cite{bruin}.} \label{Fig1C}
\end{figure}

The next example is HF metal $\rm Sr_3Ru_2O_7$ having universal
scattering rate related to the linear-temperature resistivity
\cite{bruin}, see Fig. \ref{Sc1}. This metal represents a useful
example because of numerous experimental measurements taken on
{{it}}, see e.g. \cite{bruin,pnas,sc9,physc}. {{This HF metal
is}} tuned to its quantum critical line under the application of
magnetic field $B_{c1}\leq B\leq B_{c2}$, as it is shown  Fig.
\ref{Fig1C} (a). The ordered nematic phase highlighted by the
horizontal lines is to remove the entropy excess $S(T\to0)\to
S_0$ existing in the absence of an ordered phase, therefore, an
ordered phase captures the FC state at $T\to0$
\cite{phys_rep,quasi,book}. At relatively low temperatures
$S(T)\simeq S_0$ and becomes larger than that at LFL state and
the second order phase transition converts to first one at the
tricritical points $T^1_{\rm tr}$ and $T^2_{\rm tr}$, as it is
shown in Fig. \ref{Fig1C} (a). Within the nematic phase at
$T\leq T_D$ and at $T_D\leq T$ the system exhibit the NFL
behavior with $\rho(T)\propto T$, as it is shown in Figs.
\ref{Fig1C} (a), (b). We note that at $T_D\geq T$
$\rho(T)\propto T$, but the Planckian limit does not take place,
for $d\rho(T\geq T_D)/dT > d\rho(T\leq T_D)/dT$, see Figs.
\ref{Fig1C} (b). Thus, despite the $\rho(T)\propto T$, the
universal scattering rate does not occur. This fact shows that
the Planckian limit, quantum critical behavior and the
$T$-linear resistivity are not directly related.

\begin{figure} [! ht]
\begin{center}
\includegraphics [width=0.47\textwidth] {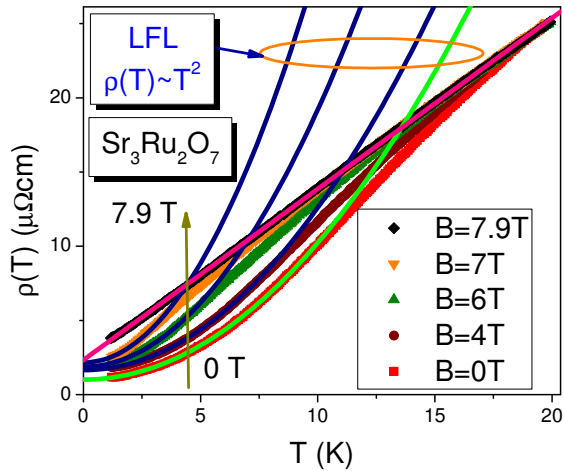}
\end {center}
\caption{(color online). The resistivity of the HF metal $\rm
Sr_3Ru_2O_7$ under the application of magnetic fields
\cite{bruin} shown at right bottom corner. The solid curves are
fits of $\rho\propto T^2$ to the low-temperature data. At
elevated temperatures the resistivity becomes $\rho\propto T$,
while at growing magnetic fields $B\to B_{c1}$ the LFL behavior
vanishes at the critical field $B_{c1}=7.9$ T, see Fig.
\ref{Fig1C} (a).} \label{Fig2}
\end{figure}

Fig. \ref{Fig2} shows the temperature dependence $\rho(T)$ of
$\rm Sr_3Ru_2O_7$ in magnetic fields.  As magnetic field $B\to
B_{c1}$, the temperature range with the LFL behavior
characterized by $\rho(T)\propto T^2$ shrinks, and at
$B=B_{c1}\simeq 7.9$ T, the resistivity $\rho(T)\propto T$ over
the whole measurement range; at $B\geq B_{c2}$ the
magnetoresistivity exhibits a small negative magnetoresistance
and the LFL behavior at low temperatures \cite{bruin}, see Fig.
\ref{Fig1C} (a). We note that such a Planckian metal does
exhibit the Planckian limit in its LFL state, and even it does
not demonstrate the limit at $T<T_D$, as we have seen above. All
these experimental observations are in accordance with general
behavior of strongly correlated Fermi systems and can be
explained {{within the FC theory framework}}
\cite{book,phys_rep}.

Thus, the fundamental picture outlined by Eq. \eqref{vf} is
strongly supported by measurements of the resistivity on $\rm
Sr_3Ru_2O_7$ for wide range of temperatures.  At $T\geq 100$ K
the resistivity becomes again the $T$-linear at all applied
magnetic fields, as it does at low temperatures and under the
application of magnetic fields $B_{c1}\geq B\geq B_{c2}$, see
Fig. \ref{Fig1C} (a). {{In the latter case, the coefficient $A$
is lower than that seen at high temperatures \cite{bruin}. This
is because the coefficient $A$ is the composition of two
contributions coming from the transverse zero sound and the FC
state, see Fig. \ref{Fig1C} (b).}} If we subtract the FC
contribution, $A$ becomes approximately the same at $T\geq T_D$
and at $T\geq 100$ K \cite{quasi}. {{Thus, similar strongly
correlated compounds exhibit the same behavior}} of the
resistivity at both quantum critical regime and high temperature
one, allowing us to expect that the same physics governs the
$T$-linear resistivity of the strongly correlated Fermi systems
and of conventional metals.

We note that there can be another mechanism supporting the
$T$-linear dependence even at $T<T_D$, {{which lifts the
constancy of $\tau$ regardless the presence}} of $T$-linear
dependence of the resistivity \cite{khod2012,quasi}. The
mechanism comes from flat bands that is formed by the FC state,
and contributes both to the linear dependence of the resistivity
and to the {\it residual} resistivity $\rho_0$, see Eq.
\eqref{res}. {{Note that these observations}} are in good
agreement with experimental data \cite{khod2012,quasi}. {{The
important point here is that under the application of magnetic
field the system in question transits from NFL behavior to LFL
one and both flat bands and the FC state are destroyed
\cite{phys_rep,book}, see the $T-B$ phase diagrams in Figs.
\ref{Fig2PD} and \ref{Fig1C} (a).}} Therefore resistivity
$\rho(T)\propto T^2$, magnetoresistance becomes negative, while
the residual resistivity $\rho_0$ {{decreases abruptly}}
\cite{khod2012,quasi,book}. Such a behavior is in accordance
with experimental data, see e. g. the case of the HF metals $\rm
CeCoIn_5$ \cite{pag1} and $\rm Sr_3Ru_2O_7$ \cite{bruin} that
also demonstrate the universal scattering rate, see Fig.
\ref{Sc1}.

{{We mention here similar recent experimental observations of
the linear resistivity $\rho(T)\propto T$ at low temperatures,
that relates the slope of the linear-T-dependent resistivity to
the London penetration depth $\lambda_0$ indicating a universal
scaling property}}
\begin{equation}\label{hzo}
\frac{d\rho}{dT} \propto \lambda_0^2
\end{equation}
for a large number of high-$T_c$ superconductors \cite{lin}.
This scaling relation spans several orders of magnitude in
$\lambda_0$, attesting to the robustness of the empirical law
\eqref{hzo}. {{This law, in turn, can be explained within the
framework of the FC theory \cite{ph_scr} rather than
addressing}} the quantum diffusion that assumed to be the origin
of this scaling relation \cite{lin}. It is important to note
that the FC theory presented here is insensitive to and
transcends the microscopic, non-universal features of the
substances under study. This is attributed to the fact that the
FC state is protected by its topological structure and therefore
represents a new class of Fermi liquids \cite{vol,book}.  In
particular, consideration of the specific crystalline structure
of a compound, its anisotropy, its defect composition, etc.\ do
not change our predictions qualitatively. {{In other words, the
fermion condensation of charge carriers in the vast body of
considered strongly correlated compounds,}} engendered by a
quantum phase transition, is indeed the primary physical
mechanism responsible for their observable universal scaling
properties. This mechanism can be extended to a broad set of
substances with a very different microscopic characteristics, as
discussed in details in Refs.~\cite{phys_rep,book}.

{{In summary, we have explained recent challenging experimental
observations that the scattering rate 1$/\tau$ of charge
carriers collected on high $T_c$ superconductors, graphene,
heavy fermion and conventional metals exhibits the universal
behavior \cite{bruin,legr,cao,nakaj} generated by the
quasiclassical properties of above strongly correlated
materials. While the Planckian limit may occur accidentally: It
is highly improbable that it would be realized in conventional
metals, which, obviously, cannot be recognized as Planckian
metals with quantum criticality at high or low temperatures.
Finally, the fact that we observe the same universal behavior of
the scattering rate in microscopically different strongly
correlated compounds like HTSC, HF and conventional metals,
suggests that some general theory is needed to explain the above
body of materials and their behavior in the uniform manner. We
may conclude that the FC theory is the suitable candidate.}}

We thank V. A. Khodel for stimulating discussions. This work was
partly supported by U.S. DOE, Division of Chemical Sciences,
Office of Basic Energy Sciences, Office of Energy Research.

\end{document}